\documentclass [epj] {svjour}
\usepackage{graphics}
\usepackage{color}
\usepackage{amsmath,amssymb,bm}
\usepackage{float}
\bmdefine\bfpihat{\hat{\pi}}
\bmdefine\bfrhohat{\hat{\rho}}
\bmdefine\bftimes{\times}

\newcommand{\Psibar}{\ensuremath{{\overline \Psi} }}
\newcommand{\PsiN}{\ensuremath{{\bm{\Psi}_{_N} }}}	   
\newcommand{\PsibarN}{\ensuremath{{\bm{\overline \Psi}}_{_N} }}
\newcommand{\PsiDeltaupmubar}{\ensuremath{{\bm{\overline \Psi}}^{\mu} _{_{\! \Delta}}}}
\newcommand{\gmunudo}{\ensuremath{g_{_{ \scriptstyle \mu \nu }}}}
\def\Lkappa{\Large \mbox{$\kappa  $} \normalsize}

\newcommand{\hatrhomoinsmu}
           {\ensuremath{{\hat{{\rho}}^{\, - ^{\scriptstyle \mu} } }}}
\newcommand{\hatrhozeromu}
           {\ensuremath{{\hat{{\rho}}^{\, 0 ^{\scriptstyle \mu} } }}}
\newcommand{\hatrhoplusmu}
           {\ensuremath{{\hat{{\rho}}^{\, + ^{\scriptstyle \mu} } }}}

\newcommand{\hatrhomoinsmuinf}
           {\ensuremath{{\hat{{\rho}}^{\, -  } }_{\scriptstyle \mu} }}
\newcommand{\hatrhozeromuinf}
           {\ensuremath{{\hat{{\rho}}^{\, 0  } }_{\scriptstyle \mu} }}
\newcommand{\hatrhoplusmuinf}
           {\ensuremath{{\hat{{\rho}}^{\, +  } }_{\scriptstyle \mu} }}

\newcommand{\hatkmoins}{\ensuremath{{\widehat{{K}}^{-} }}}
\newcommand{\hatkplus}{\ensuremath{{\widehat{{K}}^{+} }}}

\newcommand{\hatkzero}{\ensuremath{{\widehat{{K}}^{0} }}}
\newcommand{\hatkbarzero}{\ensuremath{\widehat{{\overline{K}}}^{_{_{\,\scriptstyle 0}}}} }

\newcommand{\sslashq}[1]
           {\mbox{${#1}\hspace{-0.21cm}/$}}

\newcommand{\hatsigmaplus}{\ensuremath{\widehat{\Sigma} ^{+}}}
\newcommand{\hatsigmamoins}{\ensuremath{\widehat{\Sigma}^{-}}}
\newcommand{\hatsigmazero}{\ensuremath{\widehat{\Sigma}^{0}}}
\newcommand{\hatsigmabarplus}{\ensuremath{\widehat{\overline{\Sigma}} ^{_{\, +}}}}
\newcommand{\hatsigmabarmoins}{\ensuremath{\widehat{\overline{\Sigma}} ^{_{\, -}}}}
\newcommand{\hatsigmabarzero}{\ensuremath{\widehat{\overline{\Sigma}} ^{_{\, 0}}}}

\def\hatlphi{\Large \mbox{$ \hat{\phi} $} \normalsize}

\begin{document}
\title{ Regge description of two pseudoscalar meson production in antiproton-proton annihilation}

\author{ Jacques Van de Wiele\inst{1} and Saro Ong\inst{1,2} }

\institute{ Institut de Physique Nucl\'eaire, IN2P3-CNRS, Universit\'e de
Paris-Sud, 91406 Orsay Cedex, France.
\and Universite de Picardie Jules Verne, F-80000 Amiens, France}

\date{Received: date / Revised version: date }

\abstract{A Regge-inspired model is used to discuss the hard exclusive  two-body hadronic reactions \\ 
 $( \bar{p} p \rightarrow \pi^- \pi^+~,~\pi^0 \pi^0~,~K^- K^+ ~,~{\overline {K}}^{_0} K^0)$ 
for the FAIR facility project at GSI with the Panda detector. The comparison between the differential cross sections predictions and the available data is shown to determine the values of the few parameters of the model. }

\titlerunning{Regge description of two pseudoscalar meson production in antiproton-proton annihilation}
\authorrunning{J. Van de Wiele et al.}
\maketitle
\section{  Introduction }
\label{intro}
The complex angular momentum method introduced by Regge [1] has been copiously used in particle physics. The Regge approach makes a non-trivial connection between particles, resonance spectra and high energy scattering amplitude behaviour.
In the early 1970's, there was considerable interest in two-body hadronic reactions,
both from a theoretical [2,3,4,5] and an experimental [6,7,8] view point. An important compilation of the data in the high energy regime were interpreted in terms of the Regge trajectories albeit with a limited succes.\\
Later on, the Regge poles in connection with QCD [9] and the phenomenology based on the Pomeron hypothesis [10,11] led to a predictive approach in hadron-hadron scattering in the asymptotic high energy regime with a small number of parameters.\\

The planned FAIR facility project at GSI together with the Panda detector [12] will allow one to reexamine the two-body
hadronic reactions in antiproton-proton annihilation with unprecedented high accuracy.\\

In principle, the different observables for exclusive two-body hadronic reactions could conceivably be calculated from QCD, the fundamental quantum field theory of strong interaction. In practice, such observables cannot be predicted in a simple way with only quark and gluon degrees of freedom. A huge number of diagrams has to be calculated as well as the contribution from
pinch singularities which are related to the independent multiple scattering mechanism [3].\\

As said above, the Regge phenomenology is an alternative and economical way to fit the available data of two-body hadronic reactions. The parametrization of the amplitudes in term of Regge trajectories is succesfully investigated in photoproduction and electroproduction of mesons and kaons [13,14,15].\\

It is well established that the Regge description is valid only for high energy (large $s$) but low negative momentum transfert ($-t \ll s$), where $s$, $t$ and $u$ are the usual Mandelstam variables of the process.
The linear trajectories are found to approach negative constants for large negative $t$ [16,17] and the forward and backward Regge regimes must be joined smoothly onto the fixed large angle perturbative QCD behaviour. This extrapolation of the Regge model to the larger momentum transfer region is based on the saturating Regge trajectories
($\alpha(t) \rightarrow -1$ when $t \rightarrow -\infty$ )
 and the introduction of the hadronic form factors in order to respect the dimensional counting rules [18,19].\\

In this paper, we investigate two pseudoscalar meson production in antiproton-proton annihilation. 
A phenome- \linebreak nological model based on the Regge trajectories exchanged in the $t$-channel and the vector meson in the $s$-channel is developed. An elegant 
procedure to connect the Regge trajectories in the two ranges of low and large negative
momentum transfer $t$ is discussed in detail and the comparaison with the available data for two pseudoscalar meson production is done.

\section{ Model for two body hadronic reactions}
\label{sec:1}
An attentive inspection of the two-body hadronic reactions in antiproton-proton annihilation data at $\sqrt{s} > 3$ GeV  shows a lack of theoretical investigations in both charged 
and neutral pseudoscalar meson pair production [7,8,20]. Most of the existing studies for 
$\sqrt{s} < 2$ GeV use the effective lagrangien formalism with the nucleon and $\Delta$-resonance exchange picture.\\
 In the present paper, we perform a global analysis of the world data
of two pseudoscalar meson production, using the Regge formalism in the high energy regime at invariant energy above 3 GeV.
We reggeize the nucleon and $\Delta$ exchanges by appropriate Regge trajectories $\alpha(t)$
(Figs. 1-2).

\begin{figure}[H]
\begin{center}
\resizebox{0.45\textwidth}{!}{
\includegraphics{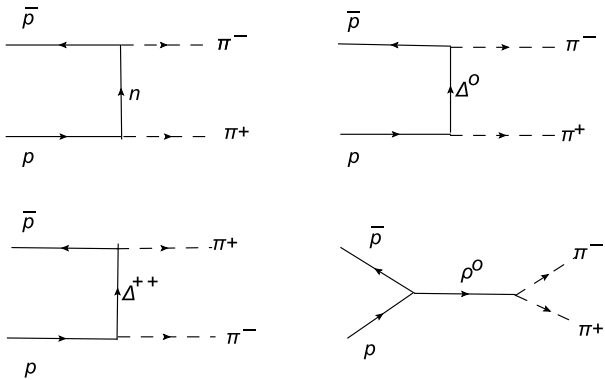}
}
\vspace*{0.0cm}
\caption{Feynman diagrams for $\bar{p} p \rightarrow \pi^- \pi^+$ }
\label{fig:1}
\end{center}
\end{figure}

A good description of the forward and backward behaviour of the differential cross sections of two-body hadronic channels was achieved in a satisfying way.
Our original idea is to perform also a systematic analysis of the data in the large angle scattering 
region at large negative momentum transfer $t$, which is the domain of predilection of perturbative QCD. To accomplish this delicate work, we assume the saturating trajectories 
for $t \rightarrow -\infty$ and the vector meson exchanged in the $s$-channel 
without reggeization to fit the available data. 
We only assume that Regge poles will control both the small $|t|$ region and
the large $|t|$ region. Futhermore the Reggeon amplitude will
dominate the Pomeron amplitude at fixed large angle and not too high energy.
The details of this approach is described below.

\begin{figure}[H]
\begin{center}
\resizebox{0.45\textwidth}{!}{
\includegraphics{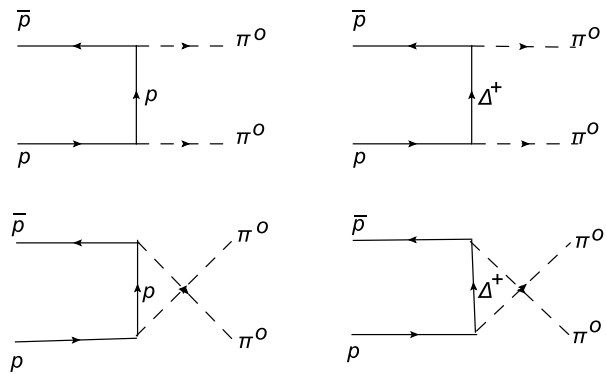}
}
\vspace*{0.0cm}
\caption{Feynman diagrams for $\bar{p} p \rightarrow \pi^0 \pi^0$ }
\label{fig:2}
\end{center}
\end{figure}

\subsection{ Regge amplitudes}

To illustrate our formalism in its technical detail, we write down explicitly the amplitudes
of the two meson production in antiproton proton annihilation. The vertex structure of theses amplitudes is kept in the usual form given by the Feynman diagrams,
using the phenomenological Lagrangians ( in appendix \ref{appa}), restricted to energies up
to the octet-baryons excitation region. Such a procedure was adopted in [13,14] for photo and electroproduction of mesons.\\

The amplitude of the nucleon-exchange diagram in the $t$-channel reads :
\begin{equation}
A_n (s,t)= C^{\pi NN}~ \bar{v}_{\bar{p}}~\gamma^5 ~\sslashq{p}_{\pi^-} {\cal P} (q_n)
\gamma^5~ \sslashq{p}_{\pi^+}~ u_p
\end{equation}
where the vertex function $C^{\pi NN}$ contains  the coupling constant and the form factor. 
The fermion propagator is the usual Feynman propagator :
$${\cal P} (q_n)={\sslashq{q}_n+M_n  \over q_n^2-M_n^2}$$
$q_n$ is the four-momentum of the exchanged nucleon.\\

And for the $\Delta$-exchange diagrams in the $t$-channel:
\begin{equation}
A_{\Delta} (s,t)= \bar{v}_{\bar{p}}~ C^{\pi \Delta N}_{\beta} ~{\cal P}^{\beta \alpha} (q_{\Delta})  ~C^{\pi \Delta N}_{\alpha}~ u_p
\end{equation}
The propagator of the $\Delta$ is the Rarita Schwinger one :
$${\cal P}^{\alpha \beta} (q_{\Delta})=-
{\sslashq{q}_{\Delta}+M_{\Delta} \over D_{\Delta}^2} \times G^{\alpha \beta}$$

$$ G^{\alpha \beta}=g^{\alpha \beta}-{1 \over 3} \gamma^{\alpha} \gamma^{\beta}
-{\gamma^{\alpha}q_{\Delta}^{\beta}-\gamma^{\beta}q_{\Delta}^{\alpha} \over 3M_{\Delta}}
-{2 q_{\Delta}^{\alpha} q_{\Delta}^{\beta} \over 3 M_{\Delta}^2}$$

$$ D_{\Delta}^2=q_{\Delta}^2-( M_{\Delta}-i \Gamma_{\Delta}/2)^2$$
The isospin coefficients in the amplitude of each diagrams in Figs. 1, 2 are displayed in table 1.

\begin{table}[H]
\caption{Isospin coefficients in the amplitude of each diagram.}
\label{tab:1}
\begin{center}
\begin{tabular}{lllll}
\hline\noalign{\smallskip}
$pn\pi^+$ &  $pp\pi^0$ & $p\Delta^0 \pi^+$ & $p\Delta^+ \pi^0$ & $ p\Delta^{++} \pi^-$\\
\noalign{\smallskip}\hline\noalign{\smallskip}
 ~~ 2  &  ~~ 1    &   ~~ 1/3   &   ~~ 2/3  &   ~~~1   \\
\noalign{\smallskip}\hline
\end{tabular}
\end{center}
\end{table}

The basic idea of the Regge approach formalism is to replace the particle exchange Feynman diagram with a mass $m$ by the exchange of certain quantum numbers. In the channels of our interest, 
this corresponds to the exchange of dominant baryon trajectories in the $t$ and $u$ channel and the usual Feynman propagator $1/(t-m^2)$ where $m$, the mass of the exchange particle, is replaced
by the Regge propagator.
\begin{equation}
\displaystyle{{1 \over t-m^2} \rightarrow {1+{\cal S} \exp(-i\pi \alpha(t)) \over
\sin(\pi \alpha(t))  \Gamma (\alpha(t)+1)} \left ( {s \over s_0} \right )^{\alpha(t)}}
\end{equation}
where $s_0=1$ GeV$^2$, $\mathcal{S} = \pm 1$ is the signature of the trajectory and 
$\alpha(t)= \alpha_0 +\alpha^{\prime} t $ is the Regge trajectory. The Regge propagator of eq. (1) is reduced to the Feynman propagator if
$t \rightarrow m^2$.
It must be noted that formula (3) is only valid for a non-degenerate trajectory of spinless particles.\\

The main ingredient in the Regge model is the trajectories defined by the spin $J$ and mass $m_J$ of the particles with a fixed $G$-parity and the relation $\alpha(m_J)=J$. For two pion production in antiproton proton annihilation, the trajectories pertinent to our approach are shown in Fig. 3 .
It is well known that Regge trajectories can be either non-degenerate or degenerate.
According to Figs. 1 and 2, the scheme adopted in our approach for two pion production in antiproton proton annihilation is discussed below.\\
In order to fit the available data, we assume a degenerate trajectory for the nucleon exchange with
$\alpha_N(t)= -0.37 + 0.98~ t $  and the Regge propagator

\begin{equation}
\displaystyle{ {\exp(-i\pi (\alpha_N(t)+1/2)) \over
\sin(\pi (\alpha_N(t)+1/2))}{\pi \alpha^{\prime}_N \over  \Gamma (\alpha_N(t)+1/2)} \left ( {s \over s_0} \right )^{\alpha_N(t)-1/2}}
\end{equation}

For the $\Delta$-exchange trajectory, a non-degenerate trajectory is used in both $t$ and $u$ channels
$\alpha_{\Delta}(t)= 0.1 + 0.93~ t $
as follows
\begin{equation}
\displaystyle{ {1- \exp(-i\pi (\alpha_{\Delta}(t)-1/2)) \over
2 \sin(\pi ( \alpha_{\Delta}(t)-1/2))}{\pi \alpha^{\prime}_{\Delta} \over   \Gamma (\alpha_{\Delta}(t)-1/2)}\hspace{-0.4mm}
 \left (\! {s \over s_0} \! \right )^{\! \alpha_{\Delta}(t)-3/2}}
\end{equation}

\subsection{The model }

For the large $-t$ and $-u$ regions, the linear Regge trajectories fail to reproduce the data. This is the hard scattering region where the reaction mechanism exhibit a different behaviour. The differential cross sections are more or less independent of $t$ and exhibit a scaling hehavior with energy as predicted by the dimensional counting rules. When extrapolated towards the large angle region
scattering, one needs to introduce a new parameter $t_{sat}$, the nominal value of $t$
where the saturing trajectory $(\alpha(t) \rightarrow -1)$ becomes effective. The original procedure
of this paper is the smooth connection between  the soft and hard mechanisms in order to always respect the concept of Regge trajectories (see fig. 3).\\
\begin{figure}
\begin{center}
\resizebox{0.40\textwidth}{!}{
\includegraphics{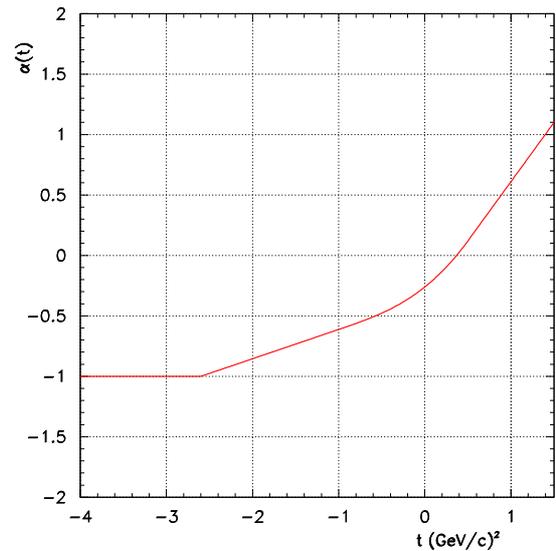}
}
\vspace*{0.0cm}
\caption{Nucleon Regge trajectory of the model in the full kinematical range of $t$ }
\label{fig:3}
\end{center}
\end{figure}

Now the question can be raised how to calculate the parameter $\alpha^{\prime}$ of the linear Regge trajectory in the transition region between the hard and soft regions.\\
Normally, $\alpha^{\prime}$ is the slope of the linear Regge trajectory in the $t>0$ region.
In the intermediate region between the asymptotic large $t<t_{sat}$ and $t<0$, we assume the
 $t$-dependency of $\alpha^{\prime}$ is determined by the slope of the straight line connecting
the corresponding point on the trajectory and those with $(\alpha(m^2) = J)$.\\

The Regge amplitude is factorized in terms of a Regge propagator and a vertex function similar to the particle exchange Feynman diagram. The vertex function contains the form factors of the exchanged particles. The monopole form is adopted in the hard scattering regime where the form factors of the outgoing particles are crucial to recover the counting rules and
the dipole forms for baryon exchanged in the soft region. A free parameter $|x_0|$ is introduced
to separate the two kinematical regions.\\ 

\noindent
For $   |x| \geq |x_{\mathrm{0}}|$, we introduce a monopole pion form factor which corresponds 
 to the hard scattering regime
\begin{eqnarray*}
 F^{\pi} _{_{M}}(x) = 
 \frac{\displaystyle \Lambda_{_{NB \pi}} ^2 - M^2 _{\pi}}
      {\displaystyle \Lambda_{_{NB \pi}} ^2 - x }
\hspace{10mm} B= N \, \text{or} \, \Delta
\end{eqnarray*}
with 
$\Lambda _{_{\scriptstyle NN \pi}} = 0.85~$  GeV,
$\Lambda _{_{\scriptstyle N \Delta \pi}} = 0.55~$ GeV
and
$x_{_{\scriptstyle \mathrm{0} N }} =x_{_{\scriptstyle \mathrm{0} \Delta }}= -1.4~ $ GeV$^2$.\\
\noindent
For $|x|  \leq |x_{\mathrm{0}}|  $
a dipole form factor is introduced both for the nucleon or the $\Delta$
\begin{eqnarray*}  
 F ^{B}_{_{D}}(x) = \bigg( 
    \frac{\displaystyle \Lambda_{_{B}} ^2 - M^2 _{_{B}} }
         {\displaystyle \Lambda_{_{B}} ^2 - x }     \bigg) ^2 
\end{eqnarray*}
\noindent
The value of the parameter $ \Lambda_{_{B}}$ is fixed
by the constraint 
at $ x = x_{\mathrm{0}}$ 
\begin{eqnarray*}
 F ^{B}_{_{D}}(x_0) =  F^{\pi} _{_{M}}(x_0) 
\end{eqnarray*}

For two charged pions channel, the $\rho$-exchange diagram in the $s$ channel is investigated
with the traditional Feynman formalism to write down the corresponding amplitude.
Guided by the previous works [13,14], we do not reggeize the $\rho$ exchange.
The vertex functions of $\rho \pi \pi$ and $\rho N N$ are given by the standard effective Lagrangian
with the coupling constants and form factors assumptions :

\begin{small}
\begin{eqnarray}\label{lagnnrho03}
\mathcal{L} _{_{ NN   \rho}} =
- K _{_{ NN \rho }} \
\bigg[
   \, \PsibarN \,   
\Big(  \gamma _{_{\scriptstyle \mu } }
      -   \frac{\Lkappa _\rho }{2 M_{_{\scriptstyle p}}}
	\, \sigma _{_{\scriptstyle \mu  \nu }}
        \, \partial ^{{\scriptstyle \nu}}    	  
\Big)  \  \bm{\tau} \cdot \bm{\hat{\rho}} ^{{\scriptstyle \mu }} \
\PsiN
\bigg]
\end{eqnarray}
\end{small}
The scalar product in isospin space is explicitly written as:
\begin{small}
\begin{eqnarray}\label{lagnnrho04}
\PsibarN \ \bm{\tau} \cdot \bm{\hat{\rho}}  ^{^{\scriptstyle \mu }}  \ 
\PsiN 
=
&&
\sqrt{2} \  \Big[
      \Psibar _p \, \hatrhoplusmu  \, \Psi _n 
   +  \Psibar _n \, \hatrhomoinsmu   \, \Psi _p
         \Big]
\nonumber \\[1mm]
&& \hspace{5mm}	 
   +  \Psibar _p \, \hatrhozeromu   \, \Psi _p
   -  \Psibar _n \, \hatrhozeromu   \, \Psi _n
\end{eqnarray}
\end{small}
The $\rho \pi \pi$ Lagrangian is taken from ref. [22]
\begin{eqnarray}\label{lagrhopipi01}
\mathcal{L}_{\rho \pi \pi} =
g_{_{\scriptstyle \rho \pi \pi}}
\big[ 
     \partial ^{^{\scriptstyle \mu}} \bfpihat \times \bfpihat
\big]   \cdot \bfrhohat _{_{\scriptstyle \mu}}   
\end{eqnarray}
which gives in terms of the physical pion fields
\begin{eqnarray}\label{lagrhopipi02}
\hspace*{-10mm}
\mathcal{L}_{\rho \pi \pi} 
 = 
i \, g_{_{\scriptstyle \rho \pi \pi}} \,
\Big\lbrace
   \hspace{4.mm}
   \big( && \hspace{-2.mm}
      \partial ^{^{\scriptstyle \mu}} \hat{\pi}^{{+}} \ \ \hat{\pi}^{{0}}
     -\partial ^{^{\scriptstyle \mu}} \hat{\pi}^{{0}} \ \ \hat{\pi}^{{+}} 
   \big)  \, \hatrhomoinsmuinf  
\nonumber\\[1mm]
&& \hspace{-7mm}
 + \hspace{1mm}   
   \big(
      \partial ^{^{\scriptstyle \mu}} \hat{\pi}^{{-}} \ \ \hat{\pi}^{{+}}
     -\partial ^{^{\scriptstyle \mu}} \hat{\pi}^{{+}} \ \ \hat{\pi}^{{-}} 
   \big)  \, \hatrhozeromuinf   
\nonumber\\[1mm]
&& \hspace{-7mm}
 + \hspace{1mm}   
   \big(
      \partial ^{^{\scriptstyle \mu}} \hat{\pi}^{{0}} \ \ \hat{\pi}^{{-}}
     -\partial ^{^{\scriptstyle \mu}} \hat{\pi}^{{-}} \ \ \hat{\pi}^{{0}} 
   \big)  \, \hatrhoplusmuinf 
   \hspace{2mm}  
\Big\rbrace      
\end{eqnarray}
with the values of the coupling constants:\\
$
 K _{_{ NN \rho }}={\sqrt{4 \pi*0.317 }}$,  
$\Lkappa _\rho = 6.033$,
$ g_{_{ \rho \pi \pi}}=\sqrt{4 \pi *2.88 }$.\\
The usual $\rho$ propagator without reggeization is:
$$ P_{_{F}} = 
\frac{\displaystyle g^{\mu \nu} - p^{\mu} p^{\nu}/M^2_{\rho}}
     {\displaystyle s - M^2_{\rho}+iM_{\rho} \Gamma_{\rho}}
$$     

The product of the form factors in the $s$-channel reads  

\begin{eqnarray*}
 F _\rho (s) = \lambda  _{_{\scriptstyle \rho }} \
\sqrt{ \bigg(
  \frac{  \Lambda   ^2 _{\scriptstyle \rho \pi \pi} - M^2 _{\pi} }
       {  \Lambda  ^2 _ {\scriptstyle \rho \pi \pi} - s }
        \bigg)  ^2}       
  \ \ 
  \frac{ M^2 _{\rho}}
       { s }
\end{eqnarray*}
with 
$\Lambda   _{\scriptstyle \rho \pi \pi} = 0.95$ GeV.
$\lambda _{_{\scriptstyle \rho }}$  is a normalization constant close to 1 taken as a free parameter.     
The value adopted in our calculation is $\lambda _{_{\scriptstyle \rho }}=0.9$.\\
To close this section on the model, we would like to emphasize the small number of the free parameters (only 7) with our Regge approach to fit the two pion channels. These parameters are :
$t_{sat,N}$, $t_{sat,\Delta}$ for saturing trajectory; $\Lambda_{NN\pi}$,$\Lambda_{NN\Delta}$ for
the form factors; $x_{0N}=x_{0\Delta}$ for the transition of the dipole to the monopole form factor 
of the exchanged particule and the outgoing particle and finally $\Lambda   _{\scriptstyle \rho \pi \pi}$,
$\lambda _{_{\scriptstyle \rho }}$ for the $\rho$ diagram contribution. We will determine the values of these parameters in the next section after a best fit with the available data.

\section{ Cross section predictions  and comparison with available data}
\label{sec:3}
Our predictions of the two pseudoscalar meson productions can be compared to the available data,
in order to determine the values of some of the free parameters of the model.
Let us emphasize the complimentary data in the charged and neutral pion channels. 
At high energy, one has only the data in the large angle scattering region for neutral pions instead of forward and backward regions in the charged channel. This leads to a strong constraints for fitting the data. \\
As can be seen in the figures 4 and 5,
the main features of the angular distribution  are reasonably
well reproduced as well as for the neutral pions as for the charged pions.
\begin{figure}[H]
\begin{center}
\resizebox{0.4\textwidth}{!}{
\includegraphics{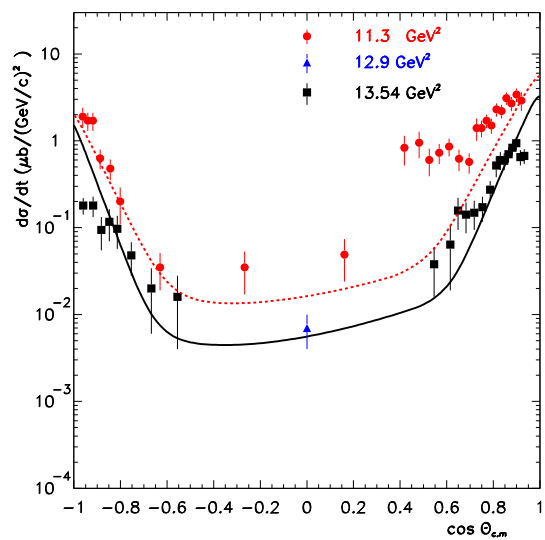}
}
\caption{Differential cross sections for $\bar{p} p \rightarrow \pi^- \pi^+$.
Data from [7,8,21], red dashed line for $s=11.3 $ GeV$^2$ and black solid line for  $s=13.54 $ GeV$^2$}
\label{fig:4}
\end{center}
\end{figure}
\begin{figure}[H]
\begin{center}
\resizebox{0.4\textwidth}{!}{
\includegraphics{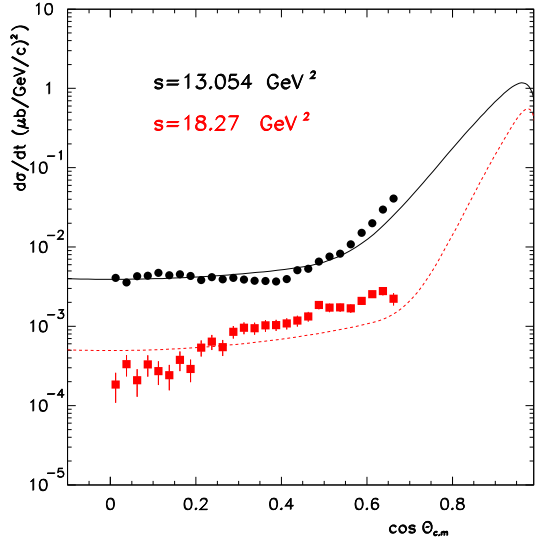}
}
\caption{Differential cross sections for $\bar{p} p \rightarrow \pi^0 \pi^0$
at $s=13.054 $ and $18.27$ GeV$^2$ . Data from [20]}
\label{fig:5}
\end{center}
\end{figure}

In this latter case, the asymmetry in the experimental cross sections
between forward (F)) ($ \cos (\theta) \sim 0.90$)  and backward (B) ($ \cos (\theta) \sim -0.90$)
angles are different at $ s = 11.3$ GeV$^2 $ ( F/B$\sim 3/2$)
and at $ s = 13.54 $ GeV$^2 $ ( F/B$\sim 1/0.2$) and this difference
as a function of the incident energy is not
reproduced by the calculation.
At $ s = 11.3 GeV^2 $, a shoulder in the angular distribution at
$ \cos (\theta) \sim 0.65$, which is missing at higher energy,
is also not reproduced by our calculation.

In Figs. 6 and 7, we display the contributions of each diagram of the two pion channels. 
\begin{figure}[H]
\begin{center}
\resizebox{0.4\textwidth}{!}{
\includegraphics{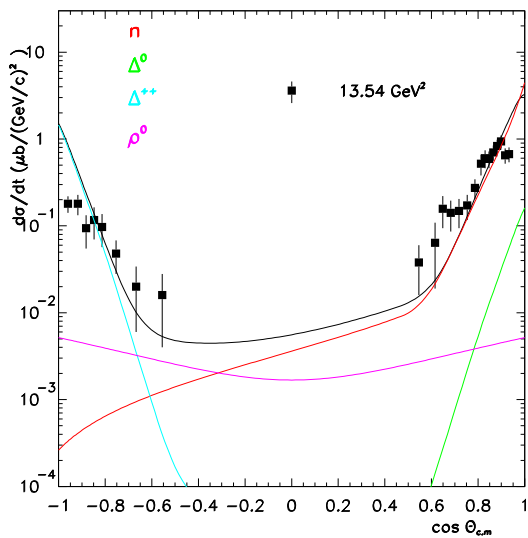}
}
\caption{Contribution of different diagrams in Fig. 1 for $\bar{p} p \rightarrow \pi^- \pi^+$
at $s=13.54 $ GeV$^2$ . Red line : nucleon, blue line :$\Delta^{++}$,
green line :$\Delta^{+}$ and pink line : $\rho$ contributions }
\label{fig:6}
\end{center}
\end{figure}
In the case of charged pions,
the forward scattering is dominated by the nucleon trajectory exchanged and the main contribution of the backward scattering is the $\Delta^{++}$ in the $u$-channel for $\pi^- \pi^+$. This is the reason of a dissymmetry only observed in the cross section for two charged pions production.
Let us emphasize that the cross section ratio between the $\Delta^{++}$ contribution at
 $\cos \theta=-1$ and the  $\Delta^{0}$ at $\cos \theta=+1$ in Fig. 6 is given by the square of the isospin coefficients in Table 1.
The contribution of the $\rho$ in the $s$-channel is small, but its interference with the others contributions could play a role to fit the data in the large angle scattering region.\\

In the case of neutral pions, the scattering is dominated by the nucleon trajectory for
$|\cos \theta| < 0.65$, while for $|\cos \theta| > 0.80$ the $\Delta$ and nucleon
contributions are the same order of magnitude.
In contrast with the charged channel, this feature could be checked with forthcoming data at
forward angle.
\begin{figure}
\begin{center}
\resizebox{0.4\textwidth}{!}{
\includegraphics{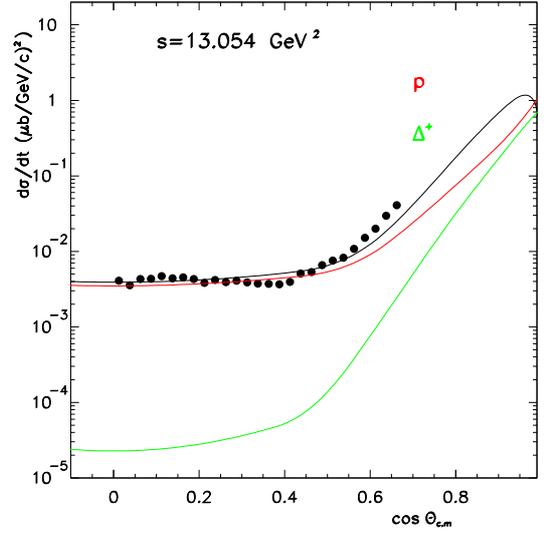}
}
\caption{Contribution of each diagram in Fig. 2 for $\bar{p} p \rightarrow \pi^0 \pi^0$
at $s=13.054 $  GeV$^2$ . Data from [20], red line : nucleon trajectory contribution; green line : the $\Delta ^+$ contribution}
\label{fig:7}
\end{center}
\end{figure}
In Fig. 8, we display the comparison between the cross section prediction of our model versus the quark interchange model with a normalisation taken from the cross channel namely the pion-proton elastic scattering [8]. The main difference is the finite value of the cross section at 
$\displaystyle{\cos \theta = \pm 1}$ and the large angle scattering $ -0.6  \leq \cos \theta \leq 0.6$ 
where our model predicts a larger value of the cross section. Indeed, the quark interchange model is not appropriate
to fit the data at forward and backward scattering and the lack of data near the $\theta =90$ degrees does not allow one to discriminate between the two models.

\begin{figure}[H]
\begin{center}
\resizebox{0.4\textwidth}{!}{
\includegraphics{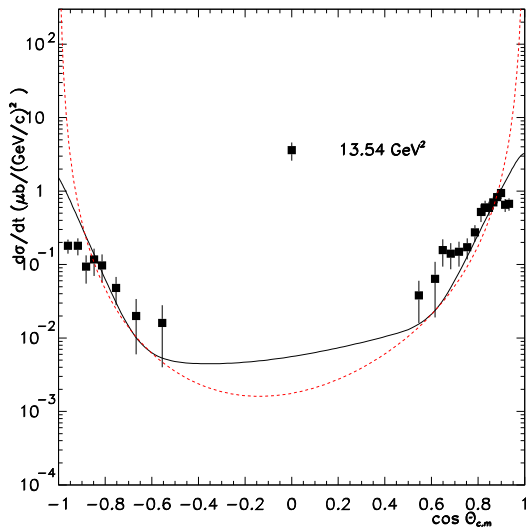}
}
\caption{Regge approach versus Quark interchange model [2] for $\bar{p} p \rightarrow \pi^- \pi^+$
at $s=13.54 $ GeV$^2$ . Red dashed line : Quark interchange model; black solid line : Regge approach model.}
\label{fig:8}
\end{center}
\end{figure}
In summary of this section, we display in the table 2, the values of different parameters of the model.
These values are obtained with a reasonable fit without standard minimization procedure,
 to leave room for future constraints from data.
\begin{table}[H]
\caption{Values of different parameters defined in the text for two pion channel}
\label{tab:2}
\begin{tabular}{lllll}
\hline\noalign{\smallskip}
$t_{sat,N}$ &$t_{sat,\Delta}$ &$\Lambda_{N N \pi}$ & $\Lambda_{N \Delta \pi}~~ x_{0,N}= x_{0,\Delta}$ & $\lambda_{\rho}~~~~\Lambda_{\rho \pi \pi}$ \\
\noalign{\smallskip}\hline\noalign{\smallskip}
-2.6 & -3 &0.85  & 0.55~~~~  ~ -1.4   &  0.9~~  0.95   \\
GeV$^2$ & GeV$^2$ & GeV &  GeV~~~~ ~ GeV$^2$  & ~~~~~~   GeV  \\
\noalign{\smallskip}\hline
\end{tabular}
\end{table}
\section{ Two kaon production}
\label{sec:4}
With the Regge approach formalism, the two kaon production channels are dominated
by the exchange mechanism of the baryons $\Lambda$ and $\Sigma$ in the $t$-channel and the $\Phi$-meson in the $s$-channel. We display in Fig. 9 the corresponding Feynman diagrams.\\
In the case of strange meson production, the exchanged baryons have to carry quantum numbers at each vertex in order to respect the strangeness conservation.
\begin{figure}[H]
\begin{center}
\resizebox{0.45\textwidth}{!}{
\includegraphics{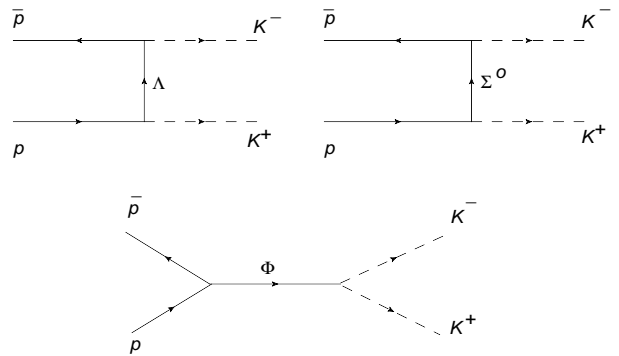}
}
\caption{Feynman diagrams for $\bar{p} p \rightarrow K^- K^+$ }
\label{fig:9}
\end{center}
\end{figure}
The Lagrangians used in our model, the coupling constants and the form factors are given in the appendix \ref{appa}.\\
Assuming the same set of free parameters of the Model
given in table 1, plus three new parameters namely\\
$\Lambda_{N \Lambda K}=\Lambda_{N\Sigma K}=1.05$ GeV for
the form factors; and finally $\Lambda   _{\scriptstyle \Phi K K}=1.3 $ GeV,
$\lambda _{_{\scriptstyle \Phi}}=1$ for the $\Phi$ diagram contribution
, we are able to fit reasonably the charged kaon channels.
As in the case of charged pions production, at $ s = 11.3$ GeV$^2 $,
the main discrepancy between the data and our caculation is located at
$ t \sim - 2$ GeV$^2$ corresponding to $ \cos (\theta) \sim 0.65$.
The agreement is better at higher energy ($ s = 13.54 $ GeV$^2 $).\\
In contrast to the two pion channels, we have no contribution from the $u$-exchanged diagram. This is an important check of the model. It seems that the data confirms this fact, but the error bars are too large to make a conclusion.
\begin{figure}
\begin{center}
\resizebox{0.4\textwidth}{!}{
\includegraphics{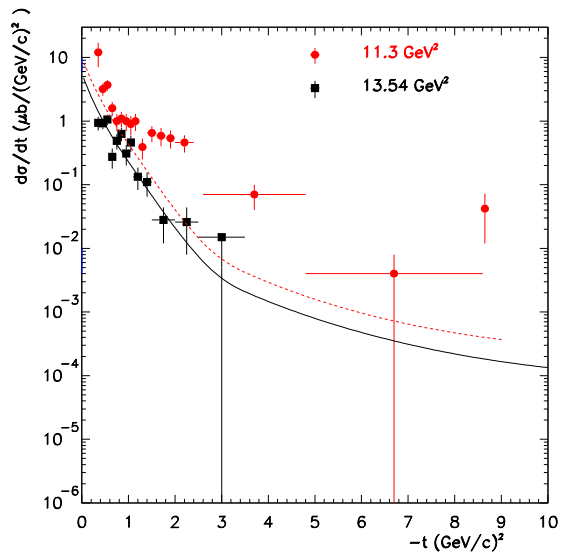}
}
\caption{Differential cross sections for $\bar{p} p \rightarrow  {K}^-  K^+$. Data from [7,8],
 red dashed line for $s=11.3$ GeV$^2$ and black solid line :  $s=13.54$ GeV$^2$ }
\label{fig:10}
\end{center}
\end{figure}
\begin{figure}[H]
\begin{center}
\resizebox{0.4\textwidth}{!}{
\includegraphics{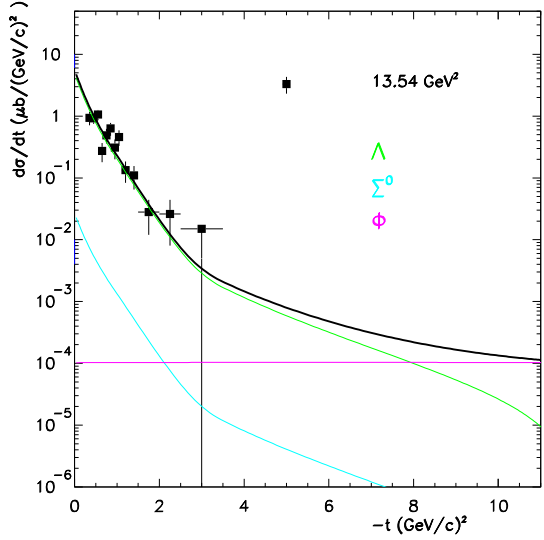}
}
\caption{Contribution of different diagrams in Fig.9 for $\bar{p} p \rightarrow {K}^-  K^+$ 
for $s=13.54$ GeV$^2$. Green line : $\Lambda$ contribution; blue line : $\Sigma^0$ contribution 
and pink line:  $\Phi$ contribution.}
\label{fig:11}
\end{center}
\end{figure}
\begin{figure}[H]
\begin{center}
\resizebox{0.45\textwidth}{!}{
\includegraphics{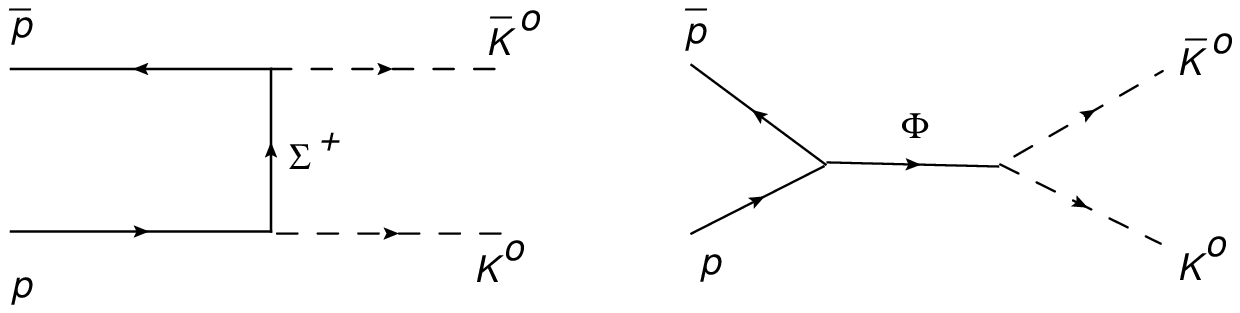}
}
\caption{Feynman diagrams for $\bar{p} p \rightarrow \bar{K}^0  K^0$ }
\label{fig:12}
\end{center}
\end{figure}
\begin{figure}[H]
\begin{center}
\resizebox{0.4\textwidth}{!}{
\includegraphics{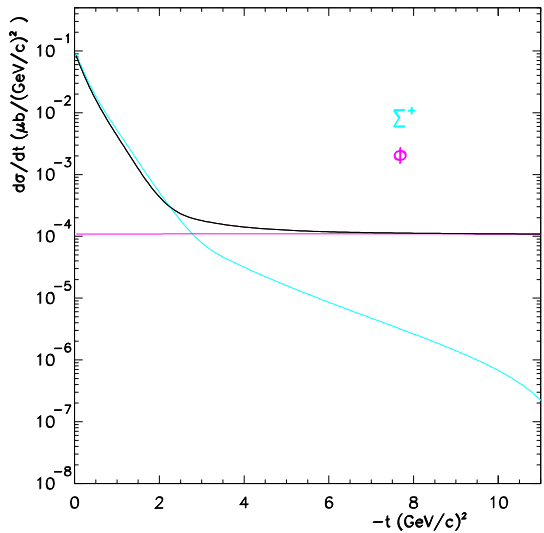}
}
\caption{Differential cross sections for $\bar{p} p \rightarrow \bar{K}^0  K^0$
with each diagram contribution of Fig. 12 at $s=13.54$ GeV$^2$.
Blue line : $\Sigma^+$ contribution; pink line : $\Phi$ contribution.}
\label{fig:13}
\end{center}
\end{figure}
\section{ Conclusions}
\label{sec:5}

In our model, there is a nice continuity between the soft and hard momentum transfer regimes.
Both are controlled by the Regge pole trajectory exchanges and the saturing trajectory at large
$-t$. With few free parameters ( 10 for two pseudoscalar meson production) 
and the Regge approach, we are able to fit reasonably well the data in antiproton proton annihilation.\\
It seems that our calculation reproduces
correctly the data for $ s \geq 13$ GeV$^2 $ and with less succes at a lower
incident energy ($ s = 11.3$ GeV$^2 $). Nevertheless, we have to make the
following remarks: i) the discrepancy with the data measured
with the same experimental set-up occurs mainly at the same angular
range for both charged pions and kaons. ii) the experimental error bars in the 
publications show large uncertainties in the data. iii) In the case of the charged pions,
the variation with the incident energy of the Forward-Backward cross 
sections should be confirmed experimentally.
Therefore, we need more data in the whole angular range 
and in a large kinetic energy domain with better statistics and small systematic errors.
 Large angle scattering data will provide constraints
on the parameters of the model.
In the hard scattering region, reliable perturbative QCD calculation of these channels would be very valuable towards a consistent conclusion.\\

The Panda detector gives an unique opportunity to measure
  these cross sections with very small statistic errors in an energy range
  5.4  GeV$^2  \leq s \leq $ 27 GeV$^2$.  For $\bar{p} p \rightarrow \pi^- \pi^+$,
the expected PANDA counting rates for one week of beam time at full luminosity $L= 2. 10^{-32}$ cm$^{-2}$ s$^{-1}$ within a $0.1$ wide $\cos \theta$
bin at $\theta=60$ degrees are displayed in table 3.\\
\begin{table}[H]
\caption{Expected counting rates of $\bar{p} p \rightarrow \pi^- \pi^+$ for different values of $s$.}
\label{tab:3}
\begin{center}
\begin{tabular}{llll}
\hline\noalign{\smallskip}
$s$ (GeV$^2$) & ~~$13.5$ & ~~~~~$16$ & ~~~~~$20$ \\
\noalign{\smallskip}\hline\noalign{\smallskip}
Event numbers & $~~10^6$  &  ~~~~~$2~10^5$   &  ~~~~~$4~10^4$   \\
\noalign{\smallskip}\hline
\end{tabular}
\end{center}
\end{table}

\section{Acknowledgments}
The authors are grateful to M. Guidal and B. Pire for useful discussions. We would like to thank the IPN PANDA group for constant encouragements and constructive remarks, in particular to R. Kunne for a
careful reading of the manuscript.
\renewcommand{\theequation}{A-\arabic{equation}}
\setcounter{equation}{0}
\begin{appendix}

\section{  Lagrangians}\label{appa}
We give the Lagrangians used in the $t$ or $u$-channels with the appropriated coupling constants.The adopted convention is that the charge associated with the field corresponds to the annihilation of the corresponding incoming particle.
\subsection{ Pion production}

We have used the
pseudo-vector $\pi NN $ lagrangian which is the one most commonly found in the literature :
\begin{small}
\begin{eqnarray}\label{lagpinnpv01}
\mathcal{L}^{^{\scriptstyle \mathrm{pv}}} _{_{ \scriptstyle  \pi N N}}
=
- K ^{^{\scriptstyle \mathrm{pv}}} _{_{ \scriptstyle  \pi N N}}
\PsibarN 
\gamma ^5 \gamma _{\mu} \, 
\bm{\tau} \cdot \big( \partial ^{\mu} \bm{\hat{\pi}} \big) \, 
\PsiN
\end{eqnarray}
\end{small}

Or explicitely, in terms of charged pion fields :

\begin{small}
\begin{eqnarray}\label{lagpinnpv05}
\hspace*{-4mm}
\mathcal{L}^{^{\scriptstyle \mathrm{pv}}} _{_{ \scriptstyle  \pi N N}}
=
&&
- K ^{^{\scriptstyle \mathrm{pv}}} _{_{ \scriptstyle  \pi N N}} \ \times
\nonumber \\
&& \hspace{-2mm}
\Big\lbrace 
\sqrt{2} \,  \Big[
      \Psibar _p \, \gamma ^5 \gamma _{\mu} \,
           \big( \partial ^{\mu}\, \hat{\pi}^{{+}}  \big) \, \Psi _n 
   +  \Psibar _n \, \gamma ^5 \gamma _{\mu} \,
          \big( \partial ^{\mu}\,  \hat{\pi}^{{-}} \big)  \, \Psi _p
         \Big]  
\nonumber \\[1mm]
&&\hspace{4mm}
+ \, \Psibar _p \, \gamma ^5 \gamma _{\mu} \, 
     \big( \partial ^{\mu}\, \hat{\pi}^{{0}} \big) \, \Psi _p	 
-    \Psibar _n \, \gamma ^5 \gamma _{\mu} \,
     \big( \partial ^{\mu}\, \hat{\pi}^{{0}} \big) \, \Psi _n
 \ \ \Big\rbrace
 \nonumber \\
\end{eqnarray}
\end{small}

with the coupling constants [26]

$K ^{^{\scriptstyle \mathrm{pv}}} _{_{ \scriptstyle  \pi N N}} =
 g _{_{ \pi N N}}/{ 2 M_{_ N} }$,
$g _{_{ \pi N N}} =\sqrt{4 \pi *12.562 }$.\\    

The $\pi N \Delta$ Lagrangian is given in a compact form similar to eq. \ref{lagpinnpv01} :

\begin{small}
\begin{eqnarray}
{\cal L }_{_{\scriptstyle \pi N \Delta }} = 
  K  _{_{ \scriptstyle  \pi N \Delta}} \
  \PsiDeltaupmubar
 \ \gmunudo 
 \ \big( \ \bm{T^+} \cdot \partial ^\nu \bm{\pi} \ \big)
 \ \PsiN + {\mathrm {h.c}}
\end{eqnarray}
\end{small}
with [26]
$K  _{_{ \scriptstyle  \pi N \Delta}} =
 g _{_{ \pi N \Delta}}/  M_{_ {\scriptstyle \pi} } $,     
$g _{_{ \pi N \Delta}} = 2.13 $        
\subsection{ Kaon production}
Following previous calculations, we assume a pseudo-scalar form for the baryon-baryon-meson lagrangian [23]

\begin{eqnarray}\label{lagnlkps02}
\hspace*{-6mm}
\mathcal{L}^{^{\scriptstyle \mathrm{ps}}} _{_{ \scriptstyle  N\Lambda K}} \!
=
 g ^{ {\scriptstyle \mathrm{ps}}} _{_{ \scriptstyle  N\Lambda K}} \,
\Big\lbrace
   \hspace{2.8mm}
      && \hspace{-1.6mm} 
        \Psibar _p \, \gamma ^5 \, \hatkplus \, \Psi _{_{\Lambda }}
      + \Psibar _{_{\Lambda }} \, \gamma ^5 \, \hatkmoins  \, \Psi _p
\nonumber \\[1mm]
&&\hspace{-4.6mm}
      + \, \Psibar _n \, \gamma ^5 \, \hatkzero  \, \Psi _{_{\Lambda }}
      + \,  \Psibar _{_{\Lambda }} \, \gamma ^5 \, \hatkbarzero  \, \Psi _p
 \ \ \Big\rbrace
\end{eqnarray}
The $N \Sigma K$ Lagrangian is written as a sum of two terms :
\begin{eqnarray}\label{lagnskps01m}
\mathcal{L}^{^{\scriptstyle \mathrm{ps}}} _{_{ \scriptstyle  N\Sigma K}}
= 
 g ^{ {\scriptstyle \mathrm{ps}}} _{_{ \scriptstyle  N\Sigma K}} \
 \Big[ \,
{\widetilde{\mathcal{L}}} ^{^{\, \scriptstyle \mathrm{ps}}} 
                          _{_{ \scriptstyle  N\Sigma K}} (I)
+
{\widetilde{\mathcal{L}}} ^{^{\, \scriptstyle \mathrm{ps}}} 
                          _{_{ \scriptstyle  N\Sigma K}}  (II)
\,  \Big]
\end{eqnarray}
with
\begin{small}
\begin{eqnarray}\label{lagnskps04m}
{\widetilde{\mathcal{L}}} ^{^{\, \scriptstyle \mathrm{ps}}} 
                          _{_{ \scriptstyle  N\Sigma K}} (I) \,
=
      && \hspace{1.8mm}
     \sqrt{2} \, \big[ \hspace{1mm}
            \hatsigmabarplus    \, \gamma ^5 \, \hatkmoins  \, \Psi _n
     \  + \ \hatsigmabarmoins   \, \gamma ^5 \, \hatkbarzero   \, \Psi _p
          \ \big]
\nonumber \\[1mm]
&&\hspace{-1mm} 
      + \  \hatsigmabarzero \, \gamma ^5 \, \hatkmoins \, \Psi _p
     \ \ - \  \hatsigmabarzero \, \gamma ^5 \, \hatkbarzero  \, \Psi _n
\end{eqnarray}
\end{small}
and
\begin{small}
\begin{eqnarray}\label{lagnskps05m}
{\widetilde{\mathcal{L}}} ^{^{\, \scriptstyle \mathrm{ps}}} 
                          _{_{ \scriptstyle  N\Sigma K}}  (II) \,
=
      && \hspace{1.8mm}
     \sqrt{2} \, \big[ \hspace{1mm}
             \Psibar _p  \, \gamma ^5 \, \hatkzero  \, \hatsigmaplus
      \ +  \ \Psibar _n  \, \gamma ^5 \, \hatkplus  \, \hatsigmamoins
          \ \big]
\nonumber \\[1mm]
&&\hspace{-1mm} 
      + \ \Psibar _p \, \gamma ^5 \, \hatkplus \, \hatsigmazero
      \ \, - \  \Psibar _n \, \gamma ^5 \, \hatkzero \, \hatsigmazero
\end{eqnarray}
\end{small}
The coupling constants  $g ^{ {\scriptstyle \mathrm{ps}}} _{_{ \scriptstyle  N\Lambda K}}$
and $g ^{ {\scriptstyle \mathrm{ps}}} _{_{ \scriptstyle  N\Sigma K}}$ are much less constrained experimentaly than  $g ^{ {\scriptstyle \mathrm{ps}}} _{_{ \scriptstyle  NN \pi}}$. The algebra associated with the two octet representations of SU(3) gives a link between these coupling 
constants [23]. We have chosen the values given in [13,27] with the upper and lower limits corresponding to 20\% breaking of the SU(3) symmetry :\\ 
\noindent
$g ^{ {\scriptstyle \mathrm{ps}}} _{_{ \scriptstyle  N\Lambda K}}=
     {\sqrt{4 \pi }}~     
\big( 
 g ^{ {\scriptstyle \mathrm{ps}}  }  _{_{ \scriptstyle  N\Lambda K \, min }} 
                  +   
 g ^{ {\scriptstyle \mathrm{ps}}  } _{_{ \scriptstyle  N\Lambda K \, max }}
 \big)/2 
$,\\[1.5mm]
\noindent
$ g ^{ {\scriptstyle \mathrm{ps}} } _{_{ \scriptstyle   N\Lambda K \, min }} 
 = - 9/2$, ~~~       
 $g ^{ {\scriptstyle \mathrm{ps}} } _{_{ \scriptstyle   N\Lambda K \, max }}     
=-3 $, \\ [1.5mm]        
\noindent
$ g ^{ {\scriptstyle \mathrm{ps}}} _{_{ \scriptstyle  N\Sigma K}}=
     {\sqrt{4 \pi }}~     
\big( 
 g ^{ {\scriptstyle \mathrm{ps}}  }  _{_{ \scriptstyle  N\Sigma K \, min }} 
                  +   
 g ^{ {\scriptstyle \mathrm{ps}}  } _{_{ \scriptstyle  N\Sigma K \, max }}
 \big) /2
$ , \\[1.5mm]
\noindent
$ g ^{ {\scriptstyle \mathrm{ps}} } _{_{ \scriptstyle   N\Sigma K \, min }} 
 = 0.9$, ~~~  
 $g ^{ {\scriptstyle \mathrm{ps}} } _{_{ \scriptstyle   N\Sigma K \, max }}     
= 1.3 $. \\      

The $NN \Phi$ Lagrangian is given in [24]:
\begin{eqnarray}\label{lagnnphi01}
\hspace*{-5mm}
\mathcal{L} _{_{ PP  \phi }} =
- K _{_{  NN \phi }} \
\bigg[
 \, \Psibar _{_{ \scriptstyle p }} \,
\Big(  \gamma _{_{\scriptstyle \mu } }
      -   \frac{\Lkappa _\phi }{2 M_{_{\scriptstyle p}}}
	\, \sigma _{_{\scriptstyle \mu  \nu }}
        \, \partial ^{^{\scriptstyle \nu}}      
\Big) \ \hatlphi ^{_{_{\, \scriptstyle \mu}}} \ \Psi _{_{\scriptstyle p}} 
\bigg]
\end{eqnarray}
with
$K _{_{  NN \phi }} = -0.90$, ~~~  
$\Lkappa _\phi =0.5  $. \\

The $KK \Phi$ Lagrangian is given in [25]:
\begin{eqnarray}\label{phikk01}
\hspace*{-7mm}
\mathcal{L}_{_{{\scriptstyle\phi} KK}} \! = - 
\frac{\displaystyle i \, g_{_{{\scriptstyle\phi} KK}}}{\sqrt{2}} \,
\Big(
 &&   \hspace{1mm}
      \hatkplus    \ \partial ^{^{\scriptstyle \mu}} \hatkmoins
\,   - \, \hatkmoins   \, \partial ^{^{\scriptstyle \mu}} \hatkplus
\nonumber \\
&&  \hspace{-2mm}
   + \, \hatkzero    \, \partial ^{^{\scriptstyle \mu}} \hatkbarzero
\,   - \, \hatkbarzero \, \partial ^{^{\scriptstyle \mu}} \hatkzero
\ \Big)   \ \hatlphi _\mu
\end{eqnarray}
\begin{eqnarray*}
g_{_{{\scriptstyle\phi} K K}} =
\left\{   \begin{array}{cc}
             6.46  &~ \text{for} \, {K}^{-} \,{K}^{+} \\[2mm]
	     6.67  & ~\text{for} \, {\overline{K}} ^{_{_{\,\scriptstyle 0}}} \, {K}^{0}
	 \end{array}
\right.	     
\end{eqnarray*}
\end{appendix}

\end{document}